\documentclass[twocolumn,amsmath,amssymb,floatfix,pra,showpacs,footinbib,superscriptaddress]{revtex4}
\usepackage{amsmath,amssymb,natbib,bm,graphicx,url,psfrag}

\usepackage{color}

\newcounter{fig}

\newcommand{\boket}[3]{\langle\, #1 \,|\, #2 \,|\, #3 \,\rangle}

\newcommand{\be}{\begin{equation}}
\newcommand{\ee}{\end{equation}}

\begin{document}
\title{Life after charge noise: recent results with transmon qubits}
\author{A.\ A.\ Houck}
\affiliation{Department of Electrical Engineering, Princeton University, Princeton, New Jersey 08544, USA}
\author{Jens Koch}
\affiliation{Departments of Physics and Applied Physics, Yale University, New Haven, Connecticut 06520, USA}
\author{M.\ H.\ Devoret}
\affiliation{Departments of Physics and Applied Physics, Yale University, New Haven, Connecticut 06520, USA}
\author{S.\ M.\ Girvin}
\affiliation{Departments of Physics and Applied Physics, Yale University, New Haven, Connecticut 06520, USA}
\author{R.\ J.\ Schoelkopf}
\affiliation{Departments of Physics and Applied Physics, Yale University, New Haven, Connecticut 06520, USA}
\date{\today}
\begin{abstract}
We review the main theoretical and experimental results for the transmon, a superconducting charge qubit derived from the Cooper pair box. The increased ratio of the Josephson to charging energy results in an exponential suppression of the transmon's sensitivity to  1/f charge noise. This has been observed experimentally and yields homogeneous broadening, negligible pure dephasing, and long coherence times of up to $3\,\mu\text{s}$. Anharmonicity of the energy spectrum is required for qubit operation, and has been proven to be sufficient in transmon devices. Transmons have been
implemented in a wide array of experiments, demonstrating consistent and reproducible results in very good agreement with theory.
\end{abstract}
\maketitle

\section{Introduction}
The idea of harnessing the power of quantum mechanics for specific computational tasks, first proposed in the early 1980s (see e.g.\ \cite{deutsch_david_quantum_1992} for an early review), has inspired physicists, engineers, and computer scientists alike. It continues to act as a prime driving force behind the ongoing research on quantum control, measurement, decoherence, and quantum information. The basic building blocks of a universal quantum computation scheme \cite{deutsch_quantum_1985} are quantum bits (qubits), which are quantum coherent two-level systems. Despite some impressive progress, the last decades have clearly established the difficulty of implementing even just a few qubits.

Nature offers only a few true two-level systems, such as spin-1/2 systems, or massless spin-1 systems (e.g.\ polarization of photons). As an alternative, sufficiently anharmonic multi-level systems can be used as effective qubits. In principle, they also offer the possibility of multi-level quantum logic \cite{muthukrishnan_multivalued_2000}. All such systems bear in common discrete energy spectra and can be understood as generalized atoms. Superconducting circuits have been established as promising systems for tunable artificial atoms: they utilize the quantum coherence of the superconducting state to minimize undesired dissipation, and employ Josephson junctions as the fundamental nonlinear and dissipationless element to obtain an anharmonic spectrum \cite{makhlin_quantum-state_2001,devoret_superconducting_2004,clarke_superconducting_2008-1}. Moreoever, fabrication of superconducting qubits benefits from the existence of well-established microfabrication techniques, spurring hope that the required scaling towards multi-qubit systems will not pose a fundamental obstacle.

Here, we review the characteristics of the transmon qubit \cite{koch_charge-insensitive_2007}, a superconducting charge qubit derived from the Cooper pair box \cite{bouchiat_quantum_1998,nakamura_coherent_1999}, with minimal sensitivity to 1/f noise. The transmon made its debut in an experiment demonstrating the photon-number dependent qubit frequency shift in the strongly dispersive limit \cite{schuster_resolving_2007}. Since then it has been successfully employed in a growing number of experiments, and has demonstrated an excellent level of agreement with theory. Coherent coupling between two transmon qubits via virtual microwave photons was reported by J.\ Majer et al.\ \cite{majer_coupling_2007}. A comprehensive verification of predicted transmon properties with high accuracy has been presented in Ref.\ \cite{schreier_suppressing_2008}, with relaxation and dephasing times in the microsecond range. The coherence times of seven different transmon devices have been analyzed and shown to be both reproducible and predictable over more than an octave in qubit frequency \cite{houck_controlling_2008}. Transmons have further been involved in recent studies of the $\sqrt{n}$ anharmonicity of the Jaynes-Cummings ladder \cite{fink_climbing_2008,bishop_nonlinear_2008}, and benchmarks of single-qubit operations \cite{chow}.

The scope of this paper is to provide a comprehensive review of the transmon basics, together with a summary of its fabrication and its overall performance as observed in recent experiments.
\begin{figure}[t]
\centering
\includegraphics[width=1.0\columnwidth]{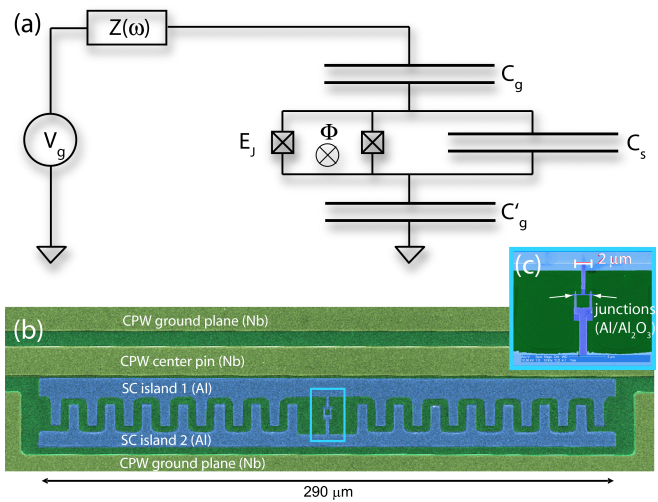}
\caption{The transmon qubit. (a) The circuit of the transmon is identical to the circuit of a differential single Cooper pair box, consisting of two superconducting islands coupled by two Josephson junctions. The coupling to ground is purely capacitive. The use of two Josephson junctions allows for tuning of the effective Josephson energy via the external magnetic flux penetrating the superconducting loop. (b) and (c) show the optical and SEM image of a transmon device positioned inside a coplanar waveguide. While the size of the junctions and the superconducting loop is very similar to CPB devices, the inter-island capacitance $C_s$ is increased drastically due to the large size of the islands and the interdigitated finger structure. This capacitance is matched by comparably large capacitances $C_g$ and $C_g'$ to the ground plane and centerpin of the transmission line resonator. \label{fig:fig1}}
\end{figure}

\section{General idea of the transmon}
Most naturally, the transmon qubit is understood as a modified version of the prototypical charge qubit, the Cooper pair box (CPB) \cite{bouchiat_quantum_1998,nakamura_coherent_1999}. As depicted in Fig.\ \ref{fig:fig1}, the transmon shares with the CPB the same underlying circuit topology. More specifically, its circuit is most closely related to the differential single Cooper pair box \cite{schneiderman_quasiparticle_2007}, which is composed of two superconducting islands and no reservoirs. Because the circuit provides no external connection between the two islands, the operator for the charge difference $n$ between the two islands has a discrete spectrum, and the superconducting phase difference $\varphi$ is only defined modulo $2\pi$, i.e., it is a compact variable for which $\varphi$ and $\varphi+2\pi z$ are considered as identical for any integer $z$ \footnote{It is important to note that these points underline the principle difference between the CPB/transmon system and the phase qubit.}. The corresponding Hamiltonian is given by
\be\label{ham}
H=4E_C(n-n_g)^2 -E_J\cos\varphi,
\ee
where the two terms describe the contribution from charging effects and Josephson tunneling, respectively. The magnitudes of these terms are set by the single-electron charging energy $E_C=e^2/2C_\Sigma$ with $C_\Sigma=C_s+(C_g^{-1}+C_g'^{-1})^{-1}$, and the Josephson energy $E_J$, which is set by the junction's normal-state conductance $G_t$ and the supeconducting gap $\Delta$ via the Ambegaokar-Baratoff relation $E_J=hG_t\Delta/8e^2$ \cite{ambegaokar_tunneling_1963}. The offset charge is denoted by $n_g$ and can be tuned by the external gate voltage.

While the CPB and transmon share the same Hamiltonian, they belong to different parameter regimes: the CPB is typically operated with $E_J\approx E_C$, and the transmon with $E_J\gg E_C$. This transmon regime is reached primarily by lowering the charging energy $E_C$. In practice, this is achieved by increasing the island sizes \footnote{The island area is increased by a factor of $\sim1000$.}, thus adding a large shunt capacitance as shown in Fig.\ \ref{fig:fig1}(b). Shunt capacitances have also been independently proposed to improve dephasing times in flux qubits by a factor of 3 \cite{you_low-decoherence_2007}, and have been implemented in phase qubits to avoid spurious resonances \cite{steffen_state_2006}.

\begin{figure}[t]
\centering
\includegraphics[width=1.0\columnwidth]{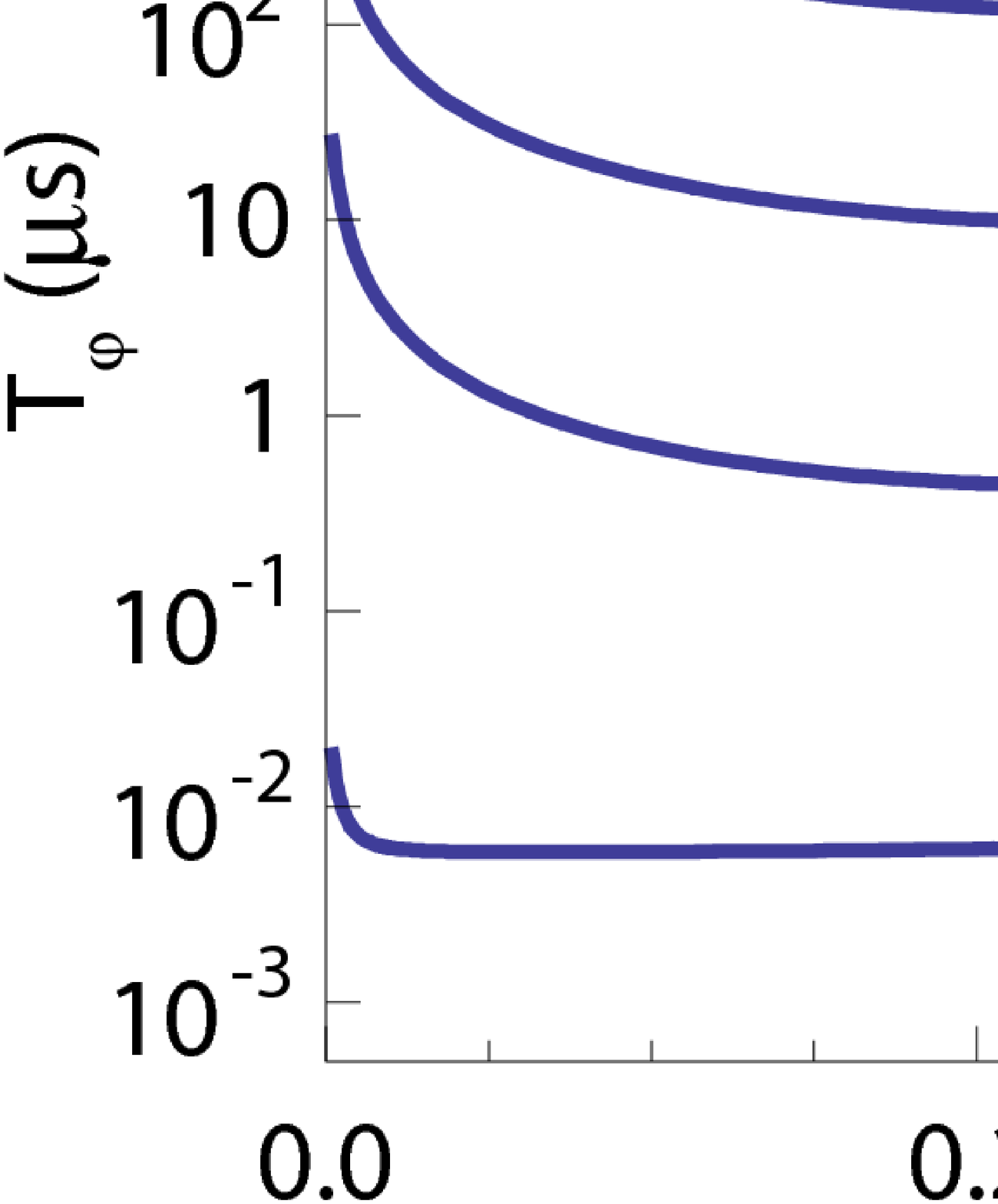}
\caption{Level spectrum and charge sensitivity of the Cooper pair box in the charge and transmon regime. (a) Eigenenergies $E_m$ (first three levels, $m=0,1,2$) of the CPB Hamiltonian \eqref{ham} as a function of the effective offset charge $n_g$ for different ratios $E_J/E_C$. All energies are given in units of the transition energy $E_{01}$ (evaluated at the degeneracy point $n_g=1/2$). The zero point of energy is chosen as the bottom of the $m=0$ level. The sequence of plots highlights the exponentially increasing flatness of energy levels and the slow loss of anharmonicity as $E_J/E_C$ is increased. (b) Order-of-magnitude estimates of dephasing times $T_\varphi$ caused by charge noise, assuming Gaussian fluctuations around a fixed offset charge, cf.\ Eq.\ \eqref{deltaomega}. Arrows on the right-hand side mark the worst-case estimates for $T_\varphi$, as determined by the total charge dispersion. The results demonstrate the exponential gain in the dephasing time due to charge noise under an increase of $E_J/E_C$. \label{fig:fig2}}
\end{figure}

Both the principle benefits and drawbacks of the transmon are evident from examining how the energy spectrum changes as one increases the ratio of Josephson energy and charging energy $E_J/E_C$ from the charging regime to the transmon regime, cf.\ Fig.\ \ref{fig:fig2}(a). In the charge regime, the spectrum of the CPB is dominated by charge parabolas with avoided crossings at the charge degeneracy points due to Josephson tunneling. If operated away from a charge degeneracy point, it is readily apparent that the qubit transition energy $E_{01}$ varies rapidly with gate charge $n_g$, thus resulting in fast dephasing due to random fluctuations in the local electrostatic potential (1/f charge noise). To some extent, this can be combatted by careful biasing of the qubit at a charge degeneracy point, where the levels are first-order insensitive to charge noise, a special operating point termed ``sweet spot" \cite{vion_manipulating_2002}. When biased at such a point, dephasing decreases; however, second-order effects of charge noise can still limit dephasing rates. As confirmed in an experiment by M.\ Metcalfe et al.\ \cite{metcalfe_measuring_2007}, second-order charge noise can indeed be identified as the clear limitation on dephasing times in the CPB charge regime. As an additional complication, drifts in the offset charge can quickly result in a departure from the sweet spot.

The main idea of the transmon is to eliminate these problems by making the charge-dependence of energy levels negligibly small. This is achieved precisely by an increase of the $E_J/E_C$ ratio, as shown in the sequence of plots in Fig.\ \ref{fig:fig2}(a). The graphs demonstrates that the energy levels become increasingly flat, i.e., independent of charge, as $E_J/E_C$ is increased. We note, however, that the energy spectrum always remains $2e$ periodic in the offset charge, illustrating that the transmon is still a charge qubit. (No such periodicity exists, e.g., for the phase qubit where $\varphi$ is not a compact variable.)
The increased flatness of levels at $E_J/E_C\gg1$ effectively generates a ``sweet spot everywhere" so that sensitivity to charge noise is suppressed to high order, and device performance is not strongly dependent on a particular bias point anymore. A detailed analysis \cite{koch_charge-insensitive_2007} shows that the suppression of charge sensitivity is exponential in the parameter $\sqrt{8E_J/E_C}$.

The effect of this exponential suppression can be recast into a statement about the dephasing induced by 1/f charge noise. A useful order-of-magnitude estimate of the dephasing time $T_\varphi$ can be obtained by using characteristic amplitudes of charge noise and evaluating the fluctuations in qubit frequency in terms of a Taylor expansion. Assuming a Gaussian process for the offset charge, the root-mean-square fluctuations in the qubit frequency are given by
\be\label{deltaomega}
\delta\omega_\text{rms}=\left[
n_\text{rms}^2 (\frac{\partial\omega_{01}}{\partial n_g})^2 + \frac{3n_\text{rms}^4}{4}(\frac{\partial^2\omega_{01}}{\partial n_g^2})^2
\right]^{1/2}.
\ee
To obtain a finite variance, the 1/f spectrum has to be cut off, and all results depend weakly (logarithmically) on the specific choice of the cutoff. Here, we use a typical value of $n_\text{rms}=0.5\cdot10^{-3}$ \cite{zorin_background_1996}. The estimate predicts a six orders of magnitude improvement in dephasing time due to charge noise $T_\varphi\approx 1/\delta\omega_\text{rms}$ by changing $E_J/E_C$ by a factor of 50, cf.\ Fig.\ \ref{fig:fig2}(b).

A worst-case estimate of the dephasing time can also be obtained by considering the maximum possible variation of the qubit transition energy, known as the charge dispersion $\epsilon_{01}$. Dephasing cannot occur faster than $T_\varphi \sim\hbar/\epsilon_{01}$, independent of the amplitude of 1/f charge noise. This scenario is depicted by arrows in Fig.\ \ref{fig:fig2}(b).  Due to the exponential suppression of level variation in the transmon regime, dephasing remains negligible even with this worst case estimate. Furthermore, a detailed analysis \cite{koch_charge-insensitive_2007} shows that the increase in $E_J/E_C$ does not heighten sensitivity to any of the other known 1/f noise mechanisms of dephasing, such as flux or critical current noise. In fact, as is evident from  Table \ref{tab1}, by operating in the transmon instead of the charge regime, one gains a factor of 2 in both the insensitivity with respect to critical current and flux noise.
\begin{table}[t]
{\centering
\begin{ruledtabular}
\begin{tabular}{lccc}
&& Transmon & CPB \\
 && $E_J/E_C=100$ & $E_J/E_C=1$ \\ 
 Noise source &$1/f$ amplitude, $A$ &$T_\varphi$ ($\mu$s) &$T_\varphi$ ($\mu$s) \\\hline
Charge & $10^{-4}$-$10^{-3}e$ \cite{zorin_background_1996} & 24,600 & \textbf{1.1}$^*$ \\ 
Flux & $10^{-6}$-$10^{-5}\Phi_0$ \cite{wellstood_low-frequency_1987,yoshihara_decoherence_2006} & 3,600$^*$ & 1,800$^*$ \\
Crit.\ current & $10^{-7}$-$10^{-6}I_0$ \cite{harlingen_decoherence_2004} & \textbf{35} & 17
\end{tabular}
\end{ruledtabular}}
$^*$ These values are evaluated at a sweet spot\\(i.e., second-order noise).
\caption{Comparison of dephasing times for the transmon and Cooper pair box qubit with $\omega_{01}/2\pi=7\,\text{GHz}$. Contributions to  $T_\varphi$ are theoretical predictions based on \cite{koch_charge-insensitive_2007}. Entries in bold face mark the dominant noise channel. For the CPB, second-order charge noise at the sweet spot limits the performance of the qubit. In contrast, for the transmon dephasing is suppressed to an extent that coherence times are limited by relaxation ($T_1$) processes only.\label{tab1}}
\end{table}

The only drawback of the $E_J/E_C$ increase is revealed by examining the level spacings in Fig.\ \ref{fig:fig2}(a). Although the charge dependence has been suppressed exponentially, the level spectrum approaches that of a pure harmonic oscillator, which would prevent the system from acting as a qubit.  However, the anharmonicity $\alpha$, determined by the difference between the fundamental qubit transition and the next higher transition frequency, decreases only slowly as function of the $E_J/E_C$ ratio, following $\alpha/\omega_{01}\sim (E_J/E_C)^{-1/2}$ \cite{koch_charge-insensitive_2007}. Thus, increasing the $E_J/E_C$ ratio from 1 to 50 can virtually eliminate the effects of charge noise, and still maintain sufficient anharmonicity to act as an effective two-level system. We illustrate this statement by considering the concrete example of a device with transition frequency $\omega_{01}/2\pi=7\,\text{GHz}$ and $E_J/E_C=100$. For these parameters, one obtains an absolute anharmonicity of $260\,\text{MHz}$ and a dephasing time due to charge noise of $T_\varphi\ge25\,\text{ms}$ (worst-case estimate).

At the large $E_J/E_C$ ratios characteristic for the transmon regime, readout of the qubit state via charge detection or measurement of the quantum capacitance \cite{duty_observation_2005} as in the CPB is not possible. In fact, the transmon does not possess any dc measurable state-dependent parameters (charge, flux, etc.). However, the transmon still exhibits a strong coupling between charge and ac voltage, rendering it an ideal candidate for an artificial atom in the circuit QED architecture \cite{blais_cavity_2004}. A full derivation of the coupling strength is given in \cite{koch_charge-insensitive_2007}, which can be intuitively understood from adding an ac component to the offset charge to describe the quantized resonator field, $n_g\to n_g^\text{dc} + C_gV_\text{rms}(a+a^\dag)/2e$. Here, $V_\text{rms}$ denotes the root-mean-square voltage of the resonator at the transmon position, and $a$, $a^\dag$ are the annihilation and creation operators for microwave photons in the relevant resonator mode. Carrying out the square in Eq.\ \eqref{ham} generates the coupling term, from which one obtains the coupling strength $g_{ij}=2eV_\text{rms}\boket{i}{n}{j}C_g^\text{eff}/C_\Sigma$ with $C_g^\text{eff}=(C_g^{-1}+{C_g'}^{-1})^{-1}$. In the transmon regime, the matrix elements are significant only between nearest-neighbor transmon levels, $i=j\pm1$, and their overall magnitude is larger by a factor of typically $3-5$ than for the CPB, due to the participation of more than one Cooper pair. In all transmons fabricated so far, the ratio $C_g/C_\Sigma$ is also $\sim3$ times larger than in the CPB, leading to a total increase in coupling strength of more than an order of magnitude.

\section{Transmon fabrication}
All transmons have thus far been fabricated in a circuit QED architecture, as the cavity offers a convenient means of reading out the qubit state. The cavity consists of a $50\,\Omega$ niobium coplanar waveguide cavity with a $4.2\,\mu\rm{m}$ gap between the center pin and ground planes. Standing waves form between transverse capacitors at either end of the waveguide, designed with a $3\,\mu\rm{m}$ spacing and a range of capacitances to vary the cavity $Q$ from $100$ to $500,000$. Cavities are patterned with photolithography and reactive ion etching of $180\,\rm{nm}$ films of sputtered niobium. Transmon qubits are patterned with electron beam lithography, and consist of two layers ($20$ and $80\,\text{nm}$ thick) of lifted-off aluminum, deposited with an angle evaporation process without venting samples to air. Samples have been fabricated on bare and oxidized silicon and sapphire substrates, with a common $Q\sim 70,000$ limit on relaxation in the devices on sapphire. Details of fabrication and processing are likely to prove important for future improvements in coherence, especially given that all transmon qubits now achieve a common limit to $T_1$.

\section{Transmon performance}
The major transmon features have been confirmed by experiments, which have been published in Refs.\ \cite{schreier_suppressing_2008} and \cite{houck_controlling_2008}. Here, we summarize these results and provide an overview of the most current transmon data. 

\begin{figure}[t]
\centering
\includegraphics[width=0.9\columnwidth]{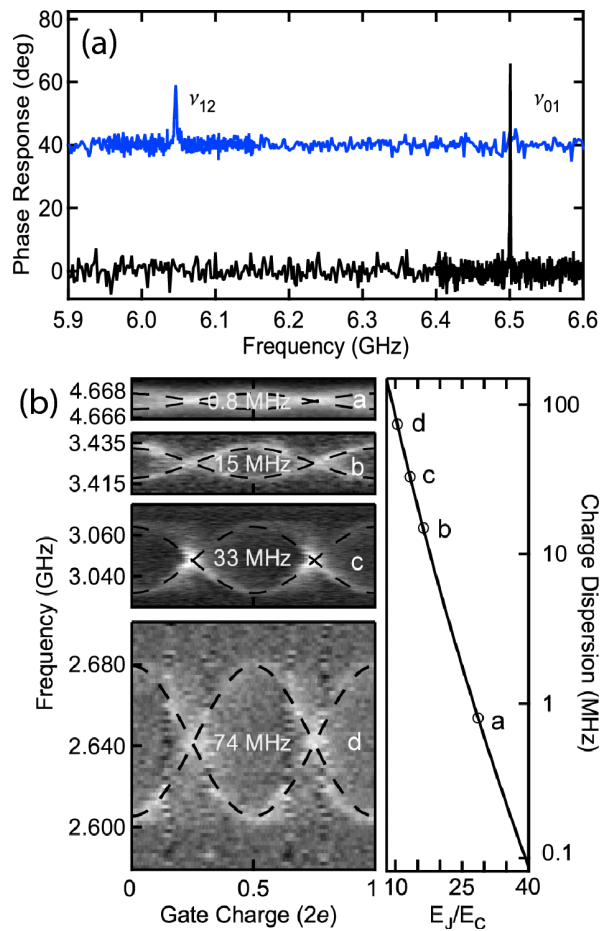}
\caption{(a) Anharmonicity of a transmon qubit.  Data presented for a transmon qubit at $E_J/E_C=40$.  The qubit transitions measured are the 01 transition in a single tone spectroscopic measurement (bottom), and the 12 transition (top, offset) while populating transmon excited state with a second drive on the 01 transition.  The second excited state of the transmon is not populated with the 12 transition at normal spectroscopy powers (bottom).  The 01 and 12 transitions are separated by $455\,\text{MHz}$; the transmon can therefore be treated as a two-level system even during fast control operations. (b) Exponential suppression of charge dispersion.  Data presented for four different values of $E_J$, where \textbf{a} $E_J/E_C=28.6$, \textbf{b} $16.3$, \textbf{c} $13.3$, and \textbf{d} $10.4$.  Spectroscopic measurements of qubit frequency while changing a gate voltage reveal the expected sinusoidal frequency bands.  The width of the band (charge dispersion) is decreased from $74$ to $0.8\,\text{MHz}$.  Two sinusoids are evident as random quasiparticle tunneling events cause the frequency curve to shift by one electron. The measured charge dispersion agrees well with the theoretical prediction (right). Reprinted from \cite{schreier_suppressing_2008}.\label{fig:fig3}}
\end{figure}
To establish that the anharmonicity is still sufficient to treat the transmon as an effective two-level system, the level spectrum of the lowest two transitions is measured by spectroscopy, as depicted in Fig.\ \ref{fig:fig3}(a). It is immediately apparent that the transitions $0\to1$ and $1\to2$ are well resolved in frequency space. The observed $455\,\text{MHz}$ anharmonicity allows for single-qubit manipulation with pulse durations of only a few nanoseconds without occupying the third level. 

The insensitivity of the transmon to charge noise, which should lead to long coherence, can directly be verified by probing the charge dependence of the transmon level spectrum. The suppression of this dependence is quantified by using the notion of charge dispersion, defined as the total variation of the qubit transition frequency (as a function of gate charge). As demonstrated by spectroscopic data in Fig.\ \ref{fig:fig3}(b), this exponential suppression agrees well with theoretical prediction and results in virtually charge-independent qubit frequencies at sufficiently large $E_J/E_C$. Instead of a single, nearly sinusoidal curve, the data displayed in Fig. \ref{fig:fig3}(b) shows two such curves with a relative displacement of half a period. This can be explained by the presence of one or several quasiparticles, which tunnel between the two islands. At low $E_J/E_C$ this phenomenon has been termed quasiparticle-poisoning \cite{aumentado_nonequilibrium_2004}, and it leads to complete dephasing of the device \cite{lutchyn_kinetics_2006}. However, the frequency shift due to such a tunneling event is bounded from above by the charge dispersion, and therefore becomes exponentially small in the transmon regime. Hence, these data confirm that the transmon is insensitive to fluctuations in local charge, including the special fluctuations due to quasiparticle poisoning.

\begin{figure}[t]
\centering
\includegraphics[width=1.0\columnwidth]{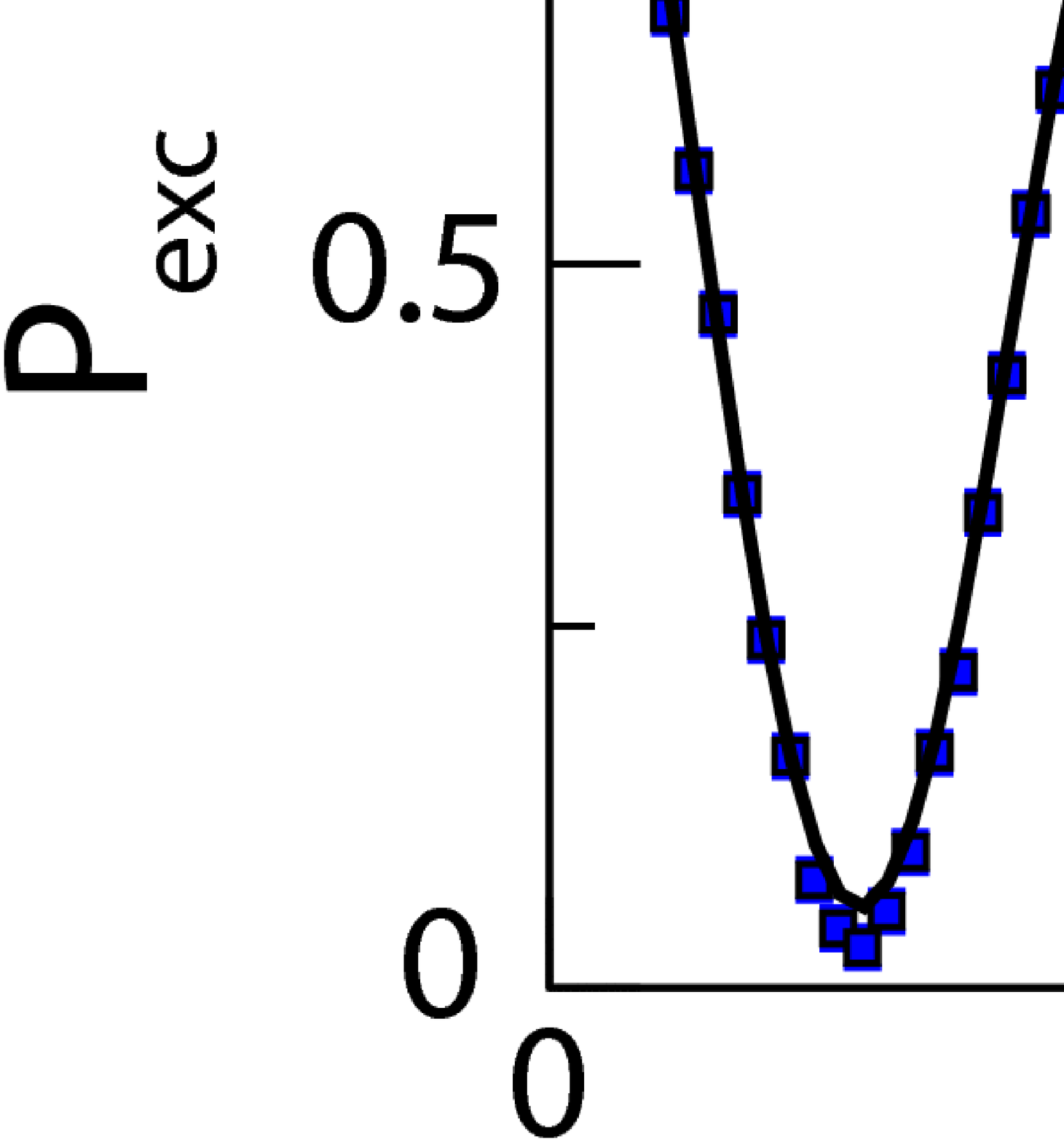}
\caption{Long coherence times in a transmon.  Data measured at the flux sweet spot for a transmon qubit with $E_C= 380\,\text{MHz}$, $E_J/E_C = 52$. (a) Relaxation from excited state.  Measurements of the occupation probability $P_\text{exc}$ of the excited qubit state while varying the time delay after a $\pi$ pulse yield an exponential decay with $T_1=1.57\pm0.04\,\mu\text{s}$. (c)  Ramsey fringes, measured by varying the time delay  between two $\pi/2$ pulses, show a long dephasing time $T_2^*=2.94\pm0.04\,\mu\text{s}$ (no echo).\label{fig:figure4}}
\end{figure}
Since charge noise was the limiting source of dephasing in charge qubits, the transmon has lead to dramatically improved dephasing times. Currently, the best $T_2^*$ for a transmon is measured at a flux sweet spot where $T_2^*=2.94\pm0.04\,\mu\text{s}$ (without any echoing) and $T_1=1.57\pm0.04\,\mu\text{s}$, see Fig.\ \ref{fig:figure4}(a) and (b) \cite{chow2}. Here, $T_2^*$ is close to identical to $2T_1$, indicating a homogeneously broadened qubit, and a  very long pure dephasing time of $T_\varphi\ge35\,\mu\text{s}$, consistent with our predictions.

\begin{figure}[t]
\centering
\includegraphics[width=1.0\columnwidth]{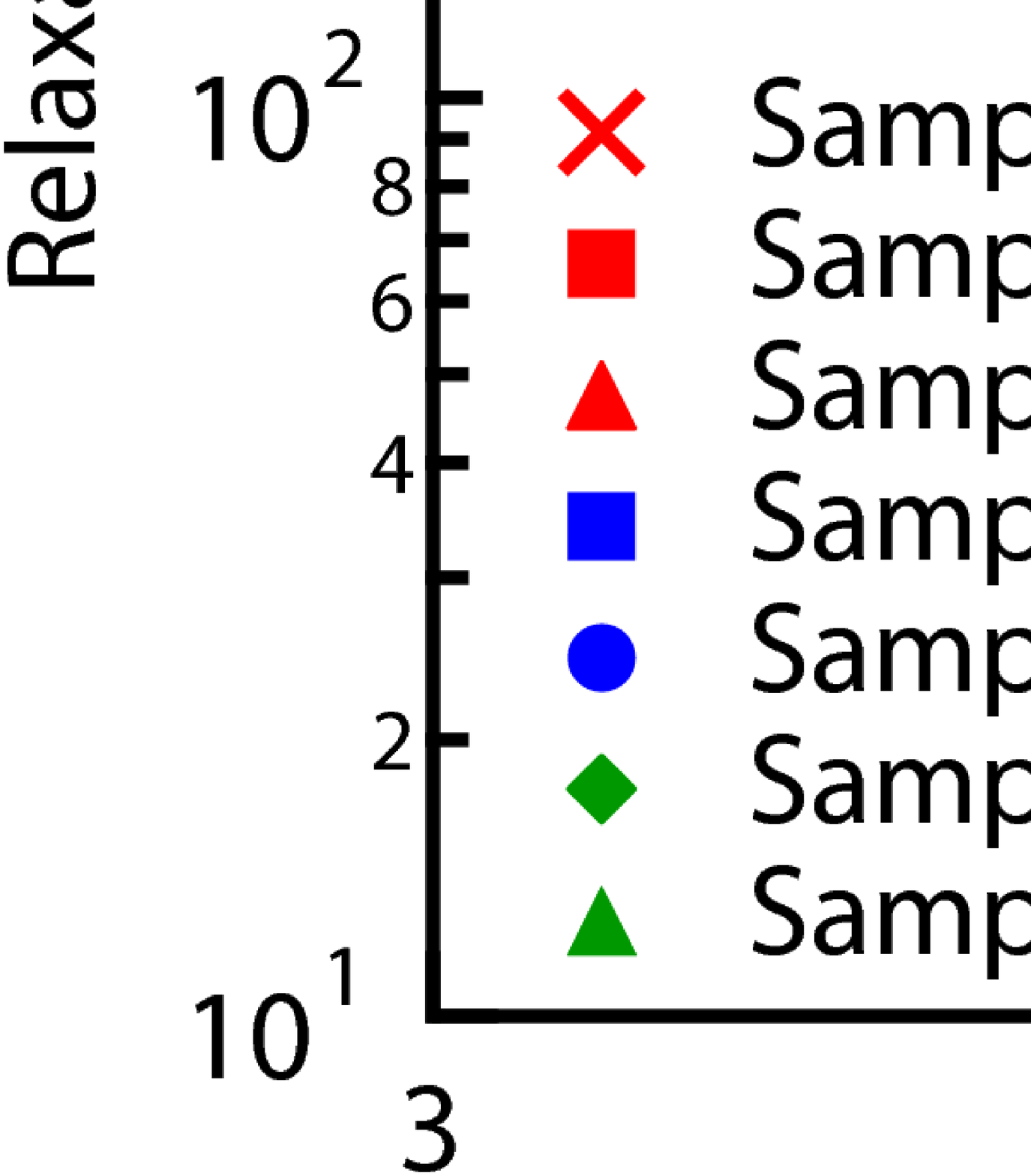}
\caption{Relaxation times for seven transmon qubits.  Predictions for qubit lifetime (colored lines) agree well with observed relaxation times (points).  Solid lines represent predictions for input side qubits located at the resonator input side (L), while dashed lines correspond to output side (R) qubits.  All sapphire qubits (blue and green) reach the same common intrinsic limits (black line), with lifetimes limited to a constant $Q\sim70,000$.  Qubit lifetimes are accurately predicted over a wide range of frequencies and more than two orders of magnitude in time. Reprinted from \cite{houck_controlling_2008}.\label{fig:figure5}}
\end{figure}

These recent results of long coherence times are to be contrasted with the performance of the first generation of transmons \cite{schuster_resolving_2007,majer_coupling_2007}, which showed shorter relaxation times of $\sim200\,\text{ns}$. The improved times shown in Fig.\ \ref{fig:figure4} are the result of a more complete understanding of relaxation, and by now have been reproduced in a number of  samples. Over more than an octave in frequency, the current limit on transmon coherence has been shown to be imposed by spontaneous emission of photons through the cavity, a process known as the Purcell effect \cite{purcell_1946}. This requires proper modeling of the cavity impedance including all higher modes of the resonator, which serves as a filter between the qubit and the evironment \cite{houck_controlling_2008}. In fact, all transmon qubits now reach the same limit on intrinsic $T_1$, as shown in Fig.\ \ref{fig:figure5}. Towards the lowest frequencies where the Purcell effect is least severe, a non-Purcell $T_1$ limitation is observed to set in, which is possibly due to dielectric loss with a $Q$ value of 70,000.

A separate issue for qubit control and usability of the qubit spectrum for operation regards the presence of coupling to unwanted degrees of freedom, such as spurious two-level systems. We have performed systematic searches of transmon spectra for such spurious resonances, which have revealed both very good agreement with theory predictions with errors as low as one part in $10^4$, and have enabled us to estimate the average number of spurious resonances in current transmon devices to be 1 per $5\,\text{GHz}$ per qubit \cite{schreier_suppressing_2008}.

\hfill
\section{Summary}
In summary, the transmon is a robust superconducting qubit for use in the circuit QED architecture.  Its primary feature is an insensitivity to 1/f charge noise, the dominant source of dephasing in other charge qubits, without any detrimental effects on the sensitivity to other known noise channels. Experiments have directly verified the predicted exponential suppression of the sensitivity to charge fluctuations by monitoring the transmon energy levels, and have confirmed the resulting gain in dephasing times with current devices reaching the $T_2=2T_1$ limit and dephasing times of up to $3\,\mu\text{s}$.
Due to the limitation of $T_2$ by relaxation, future improvements may become easier as they will focus on removing sources of dissipation. The fact that all recent transmon qubits reach a  consistent $T_1$ limit suggests that relaxation is caused by a single source, giving hope that it can be identified and eliminated in the near future.

\begin{acknowledgments}
The work reviewed in this paper was performed with the members of the Yale circuit QED collaboration: L.\ S.\ Bishop, A.\ Blais,  J.\ M.\ Chow, L.\ Frunzio, J.\ Gambetta, B.\ Johnson, J.\ Majer, J.\ A.\ Schreier, D.\ I.\ Schuster, A.\ Wallraff, and Terri M.\ Yu. This work was supported in part by Yale University via a Quantum Information and Mesoscopic Physics Fellowship (AAH, JK), by NSA under ARO contract number W911NF-05-1-0365, and the NSF under grants DMR-0653377, and DMR-0603369.
\end{acknowledgments}


\end{document}